\documentclass{ws-p8-50x6-00}

\begin{document}

\title{Chiral Symmetry and the Nucleon-Nucleon Interaction:
\\
Developing an Accurate NN Potential Based upon Chiral
Effective Field Theory
\footnote{Invited talk presented at the 7-th International Spring Seminar on Nuclear Structure 
Physics ``Challenges of Nuclear Structure'', Maiori (near Naples, Italy) May 27-31, 2001.}
}

\author{D. R. Entem\footnote{On leave from University of Salamanca, Spain.}
\, and \, R. Machleidt}
 
\address{Department of Physics, University of Idaho, Moscow, ID 83844, USA
\\
E-mails: dentem@uidaho.edu, machleid@uidaho.edu}

\maketitle

\abstracts{
We present an accurate nucleon-nucleon (NN) potential
based upon chiral effective Lagrangians.
The model includes one- and two-pion exchange contributions
up to chiral order three and contact terms (which represent
the short range force) up to order four.
Within this framework, the NN phase shifts below 300 MeV lab.\
energy and the properties of the deuteron
are reproduced with high-precision.
This chiral NN potential may serve as a reliable starting point
for testing the chiral effective field theory approach in exact
few-nucleon and microscopic nuclear many-body calculations.
}

\section{Introduction}
One of the most fundamental problems of nuclear physics is to
derive the force between two nucleons from first principles.
A great obstacle for the solution of this problem has been the fact
that the fundamental theory of strong interaction, QCD, is 
nonperturbative in the low-energy regime characteristic for
nuclear physics. The way out of this dilemma is the effective
field theory concept which recognizes different energy scales in
nature. Below the chiral symmetry breaking scale, 
$\Lambda_\chi \approx 1$ GeV,
the appropriate degrees of freedom are pions and 
nucleons interacting via a force that is governed by 
the symmetries of QCD, particularly, (broken) chiral symmetry.

The derivation of the nuclear force from chiral effective field
theory was pioneered by Weinberg\cite{Wei90}, 
Ord\'o\~nez,\cite{OK92} and van Kolck.\cite{ORK94,Kol99}
Important contributions were made by
Robilotta {\it et al.\/},\cite{RR94}
Kaiser {\it et al.\/},\cite{KBW97,KGW98,Kai99} and
Epelbaum {\it et al.\/}.\cite{EGM98}
As a result, efficient methods for deriving
the nuclear force from chiral Lagrangians have emerged.
Also, the quantitative nature of the chiral NN potential
has improved.\cite{EGM98}
Nevertheless, even the currently `best' chiral NN potentials
are too inaccurate to serve as a
reliable input for exact few-nucleon calculations or miscroscopic nuclear
many-body theory.

The time has come 
to put the chiral approach to a real test in microscopic
nuclear structure physics. Conclusive results can, however, be produced
only with a 100\% quantitative NN potential based upon chiral Lagrangians.
For this reason, we have embarked on a program to develop a NN potential
that is based upon chiral effective field theory and reproduces the
NN data with about that same quality as the 
high-precision NN potentials constructed in the 1990's.\cite{Sto94,WSS95,Mac01}

In Secs.~2 to 5, we will develop, step by step, the chiral NN potential;
and in Sec.~6 we will shown that we have achieved our goal.

\section{Effective Chiral Lagrangian}
The effective chiral $\pi N$ Lagrangian is given by
a series of terms of increasing chiral dimension,\cite{Fet00}
\begin{equation}
{\cal L}_{\pi N} 
=
{\cal L}_{\pi N}^{(1)} 
+
{\cal L}_{\pi N}^{(2)} 
+
{\cal L}_{\pi N}^{(3)} 
+ \ldots ,
\end{equation}
where the superscript refers to the number of derivatives or pion masses
(chiral dimension)
and the ellipsis denotes terms of chiral order four or higher.

{\it At lowest order}, 
the Lagrangian in its relativistic form reads
\begin{equation}
{\cal L}^{(1)}_{\pi N}  =  
 \bar{\Psi} \left(i\gamma^\mu {D}_\mu 
 - M_N
 + \frac{g_A}{2} \gamma^\mu \gamma_5 u_\mu
  \right) \Psi  
\end{equation}
with
\begin{eqnarray}
{D}_\mu & = & \partial_\mu + \Gamma_\mu 
\\
\Gamma_\mu & = & \frac12      (
                           \xi^\dagger \partial_\mu \xi
                        +  \xi \partial_\mu \xi^\dagger)
\approx 
\frac{i}{4f^2_\pi} \,
\mbox{\boldmath $\tau$} \cdot 
 ( \mbox{\boldmath $\pi$}
\times
 \partial_\mu \mbox{\boldmath $\pi$})
+ \ldots
\\
u_\mu & = & i (
                           \xi^\dagger \partial_\mu \xi
                        -  \xi \partial_\mu \xi^\dagger)  
\approx -
\frac{1}{f_\pi} \,
\mbox{\boldmath $\tau$} \cdot 
 \partial_\mu \mbox{\boldmath $\pi$}
+ \ldots
\\
U & = & \xi \xi
\\
\xi & = & e^{i\mbox{\boldmath $\tau$} \cdot 
\mbox{\boldmath $\pi$}/2f_\pi} 
\approx  1 + 
\frac{i\mbox{\boldmath $\tau$} \cdot \mbox{\boldmath $\pi$}}{2f_\pi}
-\frac{\mbox{\boldmath $\pi$}^2}{8f_\pi^2} + \ldots
\end{eqnarray}
For the parameters that occur in the first order Lagrangian,
we use $M_N=938.9187$ MeV,
$f_\pi  =  92.4$ MeV, and
$g_A  =  g_{\pi NN} \; f_\pi/M_N = 1.29$;
the latter is equivalent to 
$g_{\pi NN}^2/4\pi  =  13.67$.

We will apply
the heavy baryon (HB) formulation of chiral perturbation theory\cite{BKM95}
in which the relativistic Lagrangian is subjected
to an expansion in terms of powers of $1/M_N$ (kind of a
nonrelativistic expansion), the lowest order of which is 
\begin{eqnarray}
\widehat{\cal L}^{(1)}_{\pi N} & 
= & 
\bar{N} \left(
 i {D}_0 
 - \frac{g_A}{2} \; 
\vec \sigma \cdot \vec u
\right) N  
\\
 & 
\approx & 
\bar{N} \left[ i \partial_0 
- \frac{1}{4f_\pi^2} \;
\mbox{\boldmath $\tau$} \cdot 
 ( \mbox{\boldmath $\pi$}
\times
 \partial_0 \mbox{\boldmath $\pi$})
- \frac{g_A}{2f_\pi} \;
\mbox{\boldmath $\tau$} \cdot 
 ( \vec \sigma \cdot \vec \nabla )
\mbox{\boldmath $\pi$} \right] N + \ldots
\end{eqnarray}
In the relativistic formulation, the field operators
representing nucleons, $\Psi$, contain 
four-component Dirac spinors; 
while in the HB version, the field operators, $N$,
contain Pauli spinors; in addition,
all nucleon field operators contain
Pauli spinors describing the isospin of the nucleon.

{\it At second order},
the relativistic Lagrangian reads
\begin{equation}
{\cal L}^{(2)}_{\pi N} = \sum_{i=1}^{4} c_i \bar{\Psi} O^{(2)}_i \Psi \, .
\end{equation}
The various operators $O^{(2)}_i$ are given in Ref.\cite{Fet00}.
The fundamental rule by which this Lagrangian---as well as all the other
ones---are assembled is that they must contain {\it all\/} terms
consistent with chiral symmetry and Lorentz invariance (apart from the other trivial
symmetries) at a given chiral dimension (here: order two).
The parameters $c_i$ are known as low-enery constants (LECs)
and must be determined empirically from fits to $\pi N$ data.
We use the values determined by
B\"uttiker and Mei\ss ner,\cite{BM00}
which are (in units of GeV$^{-1}$), 
\begin{equation}
c_1 = -0.81\, , \;\;\;\;\; 
\;\;\;\;\;\;
\;\;\;\;\;\;
c_3 = -4.70\, , \;\;\;\;\; 
\;\;\;\;\;\;
\;\;\;\;\;\;
c_4 = 3.40 \, ;
\end{equation}
$c_2$ will not be needed.

The HB projected Lagrangian at order two is most conveniently broken up
into two pieces,
\begin{equation}
\widehat{\cal L}^{(2)}_{\pi N} \, = \,
\widehat{\cal L}^{(2)}_{\pi N, \, \rm fix} \, + \,
\widehat{\cal L}^{(2)}_{\pi N, \, \rm ct} \, ,
\end{equation}
with
\begin{equation}
\widehat{\cal L}^{(2)}_{\pi N, \, \rm fix}  =  
 \bar{N} \left[
\frac{1}{2M_N}\: \vec D \cdot \vec D
+ i\, \frac{g_A}{4M_N}\: \{\vec \sigma \cdot \vec D, u_0\}
 \right] N
\label{eq_L2fix}
\end{equation}
and
\begin{eqnarray}
\widehat{\cal L}^{(2)}_{\pi N, \, \rm ct}
& = & 
 \bar{N} \left[
 2\,
c_1
\, m_\pi^2\, (U+U^\dagger)
\, + \, \left( 
c_2
- \frac{g_A^2}{8M_N}\right) u_0^2
 \, + \,
c_3
\, u_\mu  u^\mu
\right.
\nonumber
\\
&& \left.
+ \, \frac{i}{2} \left( 
c_4
+ \frac{1}{4M_N} \right) 
  \vec \sigma \cdot ( \vec u \times \vec u)
 \right] N \, .
\end{eqnarray}
Note that 
$\widehat{\cal L}^{(2)}_{\pi N, \, \rm fix}$  
is created entirely from the HB expansion of the relativistic
${\cal L}^{(1)}_{\pi N}$ and thus has no free parameters (``fixed''),
while 
$\widehat{\cal L}^{(2)}_{\pi N, \, \rm ct}$
is dominated by the new $\pi N$ contact terms proportional to the
$c_i$ parameters, besides some small $1/M_N$ corrections.

{\it At third order},
the relativistic Lagrangian can be formally written as
\begin{equation}
{\cal L}^{(3)}_{\pi N} = \sum_{i=1}^{23} d_i \bar{\Psi} O^{(3)}_i \Psi \, ,
\end{equation}
with the operators, $O^{(3)}_i$, listed in Refs.\cite{Fet00,Fet98}; 
not all 23 terms are relevant to our problem.
Similar to the order two case,
the HB projected Lagrangian at order three is,
\begin{equation}
\widehat{\cal L}^{(3)}_{\pi N} \, = \,
\widehat{\cal L}^{(3)}_{\pi N, \, \rm fix} \, + \,
\widehat{\cal L}^{(3)}_{\pi N, \, \rm ct} \, ,
\end{equation}
with
\begin{eqnarray}
\widehat{\cal L}^{(3)}_{\pi N, \, \rm fix} & 
= & + \, 
   \frac{g_A}{8M_N^2} \,
\bar{N}  
\stackrel{\leftarrow}{D}
 \cdot \;
 (\vec \sigma \cdot \vec u) 
 \; \vec D\, N
\nonumber
\\
 &&
 - \, \frac{g_A}{8M_N^2} \,
\bar{N}  
\left[ ( \vec \sigma \cdot \stackrel{\leftarrow}{D} )
 \, (\vec u \cdot \vec D ) 
  + \mbox{h.c.} \right] N
\nonumber
\\
 && 
+  \, \frac{g_A}{16M_N^2} \,
\bar{N}  
  \left[ (\vec \sigma \cdot \vec u)\; \vec D^2 
  + \mbox{h.c.} \right] N \,
  + \ldots \, ,
\label{eq_L3fix}
\end{eqnarray}
the ellipsis standing for relativistic correction terms
not needed here,
and
$\widehat{\cal L}^{(3)}_{\pi N, \, \rm ct}$
given in Refs.\cite{Fet00,Fet98}; the latter reference contains also
a determination of the $d_i$ LECs.

\begin{figure}[t]
\vspace*{-1.5cm}
\hspace*{.75cm}
\psfig{figure=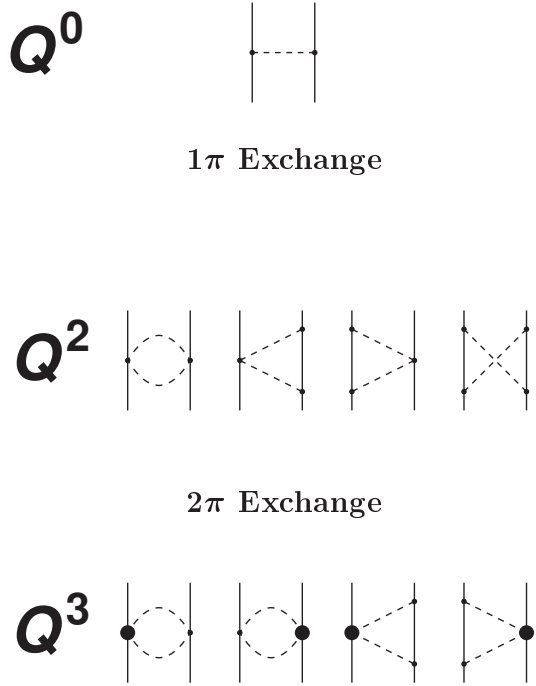,height=14.0cm}
\vspace*{-3.5cm}
\caption{The most important irreducible one- and two-pion exchange contributions to the NN
interaction up to order $Q^3$. Vertices denoted by small dots are from
$\widehat{\cal L}^{(1)}_{\pi N}$, 
while large dots refer to
$\widehat{\cal L}^{(2)}_{\pi N, \, \rm ct}$.}
\end{figure}

\begin{figure}[t]
\vspace*{-2.2cm}
%\hspace*{.5cm}
\psfig{figure=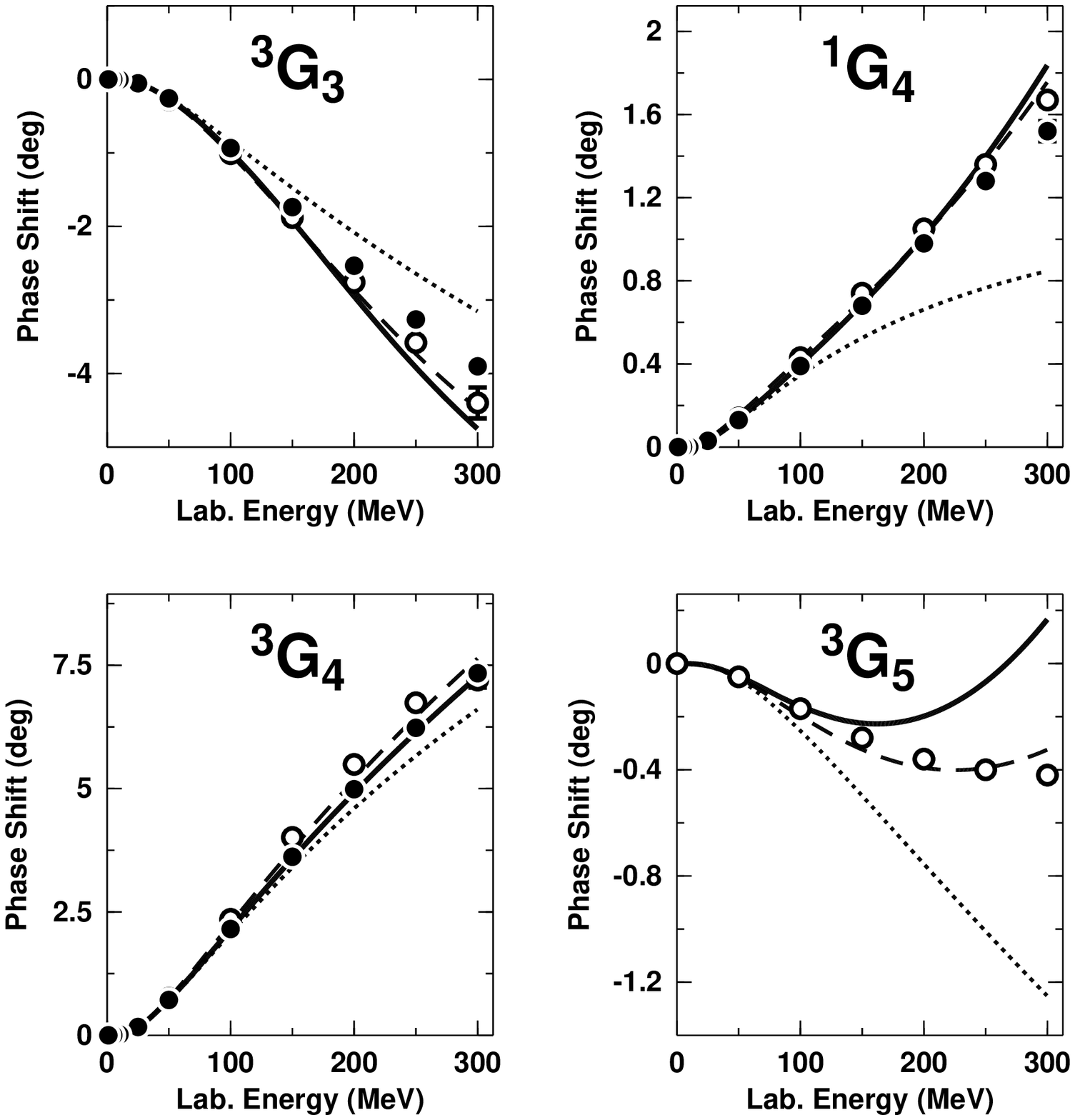,height=16.0cm}
\vspace*{-3.3cm}
\caption{$G$-wave phase shifts.
The predictions from the chiral model displayed in Fig.~1 are shown by
the solid curve and the ones from the Bonn model\protect\cite{MHE87} by the dashed curve.
The dotted curve is OPE. Solid dots represent the Nijmegen multi-energy
$np$ analysis\protect\cite{Sto93} and open circles the VPI/GWU 
analysis.\protect\cite{SM99}}
\end{figure}

\section{Pion-Exchange Diagrams and Power Counting}

The $\pi N$ Lagrangian constructed in the previous
section is the crucial ingredient for the evaluation of
the pion-exchange contributions to the NN interaction.
Since we are dealing here with a low-energy
effective theory, it is appropriate to analyze the contributions
in terms of powers of small momenta: $(Q/\Lambda_\chi)^\nu$,
where $Q$ is a generic momentum or a pion mass and $\Lambda_\chi \approx 1$ GeV
is the chiral symmetry breaking scale. 
This procedure has become known as {\it Power Counting}.
For the pion-exchange diagrams relevant to our problem, the power $\nu$ of a diagram
is determined by the simple formula
\begin{equation}
\nu = 2\times loops + \sum_j (d_j - 1) \, ,
\end{equation}
where `$loops$' denotes the number of loops in the diagram, 
$d_j$ the number of derivatives involved in vertex $j$, and the sum
runs over all vertices.

The most important irreducible one-pion exchange (OPE) and two-pion exchange (TPE)
contributions to the NN interaction
up to order $Q^3$ are shown in Fig.~1; they have been evaluated by
Kaiser {\it et al.}\cite{KBW97} using covariant perturbation theory and
dimensional regularization.
In addition to the diagrams displayed, we
take into account the relativistic corrections up to order three
as implied by
$\widehat{\cal L}^{(2)}_{\pi N, \, \rm fix}$, Eq.~(\ref{eq_L2fix}) 
and
$\widehat{\cal L}^{(3)}_{\pi N, \, \rm fix}$, Eq.~(\ref{eq_L3fix}),
where the latter contributes only to the one-pion exchange---to the order
we are working at. 

One- and two-pion exchanges are known to describe NN scattering
in peripheral partial waves. Therefore, we show in Fig.~2 (solid line)
predictions by the chiral model displayed in Fig.~1 (plus realtivistic corrections
up to order three) for the phase shifts
in $G$ waves. To provide a comparison
with conventional meson theory, we show also the predictions for $\pi + 2\pi$ exchange by
the Bonn model\cite{MHE87} (dashed line).
The dotted line represents pure one-pion exchange.
Note that our calculations with $2\pi$ models always
include also the iterated one-pion exchange.
From Fig.~2 we can conclude that, in $G$ waves (orbital angular momentum
$L=4$), there is good agreement
between the chiral and conventional $2\pi$ model as well as the empirical phase
shifts. This is also true for all partial waves with $L>4$. 

The agreement deteriorates when proceeding to lower $L$. While in $F$ waves
the agreement between the chiral model and the empirical phase shifts is
still fair, substantial discrepancies emerge in $D$ waves
as demonstrated in Fig.~3: the chiral $2\pi$ exchange is far too
attractive---a fact that has been noticed before.\cite{KBW97,KGW98}

\begin{figure}[t]
\vspace*{-8.3cm}
\hspace*{-2.0cm}
\psfig{figure=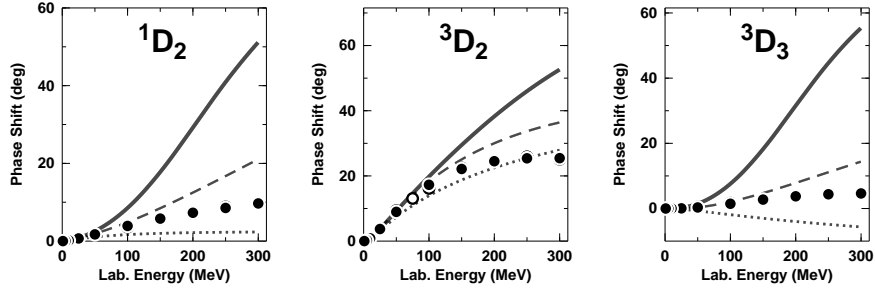,width=16.0cm}
\vspace*{-8.8cm}
\caption{$D$-wave phase shifts. Notation as in Fig.~2.}
\end{figure}

\section{Short-Range/Contact Contributions}

To control the $D$ (and lower) partial waves, we need (repulsive) short-range contributions.
In the conventional meson model, these are created by the exchange of
heavy mesons (notably, the $\omega$ meson). 
In chiral perturbation theory ($\chi$PT), heavy mesons have no place
and the short-range force is parametrized in terms of contact potentials,
which are organized by powers of $Q$.
If $Q$ is, e.~g., a momentum transfer, i.~e., $\vec Q = {\vec p}~' - \vec p$,
where $\vec p$ and ${\vec p}~'$ are the CM nucleon momenta before and after
scattering, respectively, and $\theta$ is the scattering angle, then,
for even $\nu$,
\begin{equation}
{\vec Q}^\nu \sim (\cos \theta)^m 
\hspace*{.5cm}
\mbox{with}
\hspace*{.5cm}
m\leq \frac{\nu}{2} 
\, .
\end{equation}
Partial-wave decomposition for orbital-angular momentum $L$
yields,
\begin{equation}
\int_{-1}^{+1} {\vec Q}^\nu P_L(\cos \theta) d\cos \theta \neq 0
\hspace*{.5cm}
\mbox{for}
\hspace*{.5cm}
L\leq \frac{\nu}{2} \, ,
\end{equation}
where $P_L$ is a Legendre polynominal.
The conclusion is that for non-vanishing contributions in $D$ waves ($L=2$),
$\nu=4$ is required. 
Based upon invariance considerations, there are a total of 24 contact terms
up to order $Q^4$, which we all include in our model.
 The parameters of these terms have to be natural,
but are otherwise not restricted and, thus, represent essentially free
parameters.

The ideas of $\chi$PT may suggest that, if contacts are included up to order
four, then also the $2\pi$ contribution should be calculated up to
order four. (We went up to third order in the previous section,
cf.\ Fig.~1.) We have looked into this issue and found that the 
current status of the chiral $2\pi$ exchange at order four
is a Pandora's Box.
Some contributions have been calculated, like the football diagram
with both vertices taken from
$\widehat{\cal L}^{(2)}_{\pi N, \, \rm ct}$, i.~e., both vertices proportional
to one of the LECs $c_i$. It turns out\cite{Kai01} that 
the attraction generated by
this diagram alone is
so huge that it essentially doubles
the size of the chiral $2\pi$ exchange in peripheral partial waves,
leading to very serious (irreparable?) discrepancies
with the empirical phase shifts and the predictions by conventional meson-exchange
models. Other contributions at order four are obtained by inserting
for the solid-dot vertices in Fig.~1 interactions from
$\widehat{\cal L}^{(3)}_{\pi N, \, \rm ct}$ (proportional to the LECs $d_i$). 
Consistency requires
that these diagrams are calculated together with corresponding two-loop
contributions of order four. All this is very involved and 
it will create horrific mathematical expression to just
describe the long-range NN interaction. 
At this time, nothing is known about these contributions, 
but the most optimistic prognosis would be that
the latter contributions will compensate the enormous attraction from the $c_i^2$
football. But even if this optimistic scenario were to come true,
the chiral model would no longer be practical for nuclear physics purposes.
It may then be much more reasonable to use
the `resonance saturation' argument and return to the conventional
meson models of the past, particularly, the beautiful and simple
one-boson-exchange model:\cite{Mac89} in this model the one-sigma exchange 
(the mathematical expression for which can be written in just
one line) describes the entire $2\pi$ exchange---quantitatively!

In conclusion: for the time being, the only realistic avenue towards
a quantitative NN potential based upon $\chi$PT
is a `split approach': chiral $2\pi$ up to order three,
contacts up to order four. This is our model.

\section{The NN Potential}

Since the $2\pi$ exchange diagrams, Fig.~1, are calculated using
covariant perturbation theory,\cite{KBW97} it is appropriate to
start from
 the Bethe-Salpeter (BS) equation\cite{SB51}
which reads in operator notation
\begin{equation}
{\cal T=V+V\,G\,T}
\end{equation}
with ${\cal T}$ the invariant amplitude for the two-nucleon scattering process,
${\cal V}$ the sum of all connected two-particle irreducible diagrams, and 
${\cal G}$ the relativistic two-nucleon propagator. 
The BS equation
is equivalent to a set of two equations:
\begin{eqnarray}
{\cal T}&=&\bar{V}+\bar{V}g\; {\cal T}
\label{eq_bbs}
\\
\bar{V}&=&{\cal V + V\;(G}-g)\bar{V}
\\
&\approx&
{\cal V + V_{\rm OPE}\;(G}-g){\cal V_{\rm OPE}}
\label{eq_box}
\end{eqnarray}
where the last line states the approximation we are using,
exhibiting the way we treat the $2\pi$ box diagram
($\cal V_{\rm OPE}$ is the relativistic one-pion exchange and
$\cal V$ contains all the irreducible diagrams of the type displayed in
Fig.~1, including relativistic corrections up to order three). 
This treatment avoids double counting when $\bar V$ is
iterated in the scattering equation and is also
consistent with the calculations of Ref.\cite{KBW97}.
For the relativistic three-dimensional propagator $g$, we choose
the one proposed by 
Blankenbecler and Sugar\cite{BS66} 
(BbS)\footnote{For a derivation of the BbS approach, see appendix A.1
of Ref.\cite{Mac01}.}
which has the great
practical advantage that the OPE (and the entire potential)
becomes energy-independent.
Thus, we do not need the rather elaborate formalism of unitary
transformations\cite{EGM98} to generate energy-independence of the potential.

Our full chiral NN potential $\bar V$ is defined by
\begin{equation}
\bar{V}({\vec p}~',{\vec p}) \equiv
\left\{ \begin{array}{c}
\mbox{ sum of irreducible}\\
\mbox{\boldmath $\pi + 2\pi$ contributions}
\end{array} \right\} 
+ \mbox{ contacts} \, ,
\label{eq_pot1}
\end{equation}
where the first term on the r.h.s.\ is given by 
Eq.~(\ref{eq_box}).

This potential satisfies the relativistic
BbS equation, Eq.~(\ref{eq_bbs}).
If we define now,
\begin{equation}
{V}({\vec p}~',{\vec p})
\equiv \sqrt{\frac{M_N}{E_{p'}}}\:  
\bar{V}({\vec p}~',{\vec p})\:
 \sqrt{\frac{M_N}{E_{p}}}
\approx \left(1
-\frac{p'^2+p^2}{4M_N^2}
\right)
\bar{V}({\vec p}~',{\vec p}) 
\label{eq_pot2}
\end{equation}
with $E_p\equiv \sqrt{M_N^2 + {\vec p}^2}$,
then $V$ satisfies the usual, nonrelativistic
Lippmann-Schwinger (LS) equation,
\begin{equation}
 {T}({\vec p}~',{\vec p})= {V}({\vec p}~',{\vec p})+
\int d^3p''\:
{V}({\vec p}~',{\vec p}~'')\:
\frac{M}
{{ p}^{2}-{p''}^{2}+i\epsilon}\:
{T}({\vec p}~'',{\vec p}) \, .
\label{eq_LS}
\end{equation}
Note that the correction term
$-(p'^2+p^2)/4M_N^2$ in Eq.~(\ref{eq_pot2})
is included only for OPE; for TPE it would create contributions
beyond the order to which we calculate and for
contacts it creates either existing terms or goes beyond our accuracy.

In summary, our chiral NN potential $V$ is defined by Eq.~(\ref{eq_pot2})
with $\bar V$ as given in Eq.~(\ref{eq_pot1}).
Since $V$ satisfies Eq.~(\ref{eq_LS}), it is suitable for application
in conventional, nonrelativistic nuclear structure physics.

Iteration of $V$ in the LS equation
requires cutting $V$ off for high momenta to avoid infinities.
Therefore, we regularize $V$ in the following way:
\begin{eqnarray}
V(\vec{ p}~',{\vec p})& 
\longmapsto&
V(\vec{ p}~',{\vec p})
\;\mbox{\boldmath $e$}^{-(p'/\Lambda)^{2n}}
\;\mbox{\boldmath $e$}^{-(p/\Lambda)^{2n}}
\\
&&
\approx
V(\vec{ p}~',{\vec p})
\left\{1-\left[\left(\frac{p'}{\Lambda}\right)^{2n}
+\left(\frac{p}{\Lambda}\right)^{2n}\right]+ \ldots \right\} \, ,
\end{eqnarray}
where the last equation gives an indication of the fact that
the exponential cutoff does not affect the order to which we are
calculating, but introduces contributions beyond that order.
For the contact potentials, we use cutoff masses $\Lambda$
which are partial wave dependent. One can show that in doing so
we just generalize the above regularization concept in the following sense:
\begin{equation}
V(\vec{ p}~',{\vec p}) 
\longmapsto
V(\vec{ p}~',{\vec p})
\left\{1-b_1\left[\left(\frac{p'}{\Lambda}\right)^{2n}
+\left(\frac{p}{\Lambda}\right)^{2n}\right]+ \ldots \right\} \, ,
\end{equation}
with $b_i$ of ${\cal O}(1)$.

For OPE, we use $n=4$ and $\Lambda = 0.6$ GeV; 
for TPE, $n=2$ and $\Lambda = 0.46$ GeV; and 
for the contact potentials, $n=2$ (except for contacts of order
$Q^0$ where $n=3$) and $\Lambda \approx 0.4 - 0.5$ GeV.
The number of cutoff parameters is 22 which,
together with the 24 contact parameters, results in a total
of 46 parameters for our model.
At first glance, this may sound a lot.
Note, however, that the Nijmegen phase shift
analysis\cite{Sto93} uses 40 parameters 
and that the high-precision potentials\cite{Sto94,WSS95,Mac01} 
developed in the 1990's have between 40 and 50 parameters. 
Thus, a precise fit of the NN data requires around 50 parameters,
unless a model has more predictive power than
the meson model. An important comment that has to be made about
the chiral NN potential is that it has very little predictive 
power---less than the meson model. In the light of this fact, the
number of 46 parameters is no surprise.

\section{Results for the Two-Nucleon System}

\subsection{Two-Nucleon Scattering}

\begin{figure}[t]
\vspace*{-2.0cm}
\hspace*{-1.7cm}
\psfig{figure=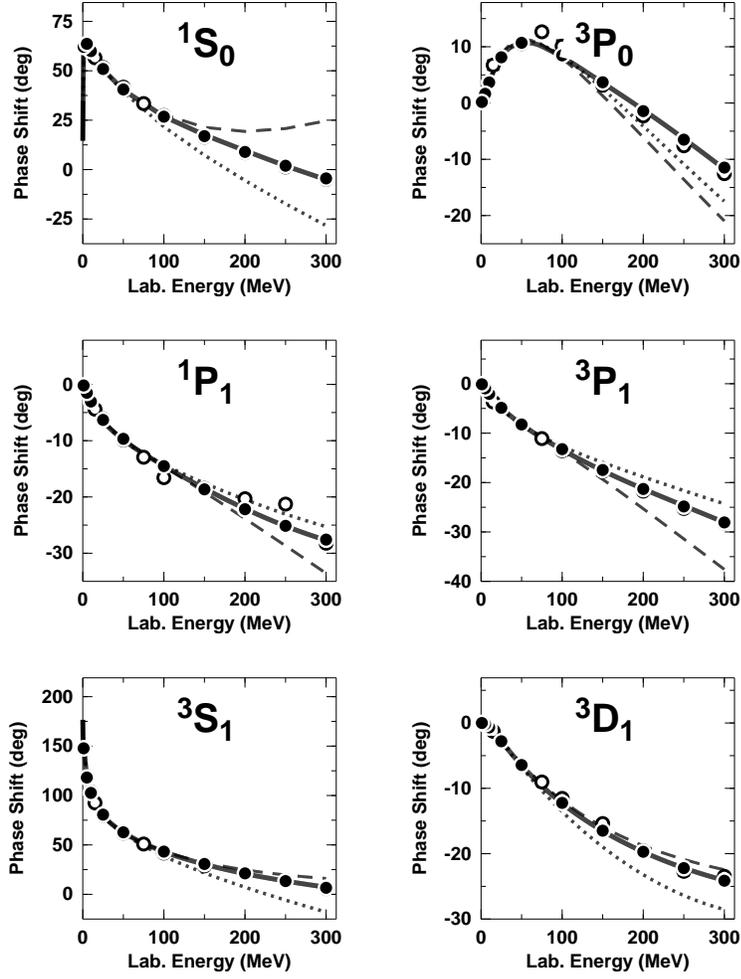,height=18.0cm}
\vspace*{-3.3cm}
\caption{Phase shifts for $J\leq 1$.
The solid line is the result from our chiral NN potential,
while the dotted and dashed lines are the predictions by two
chiral models developed by Epelbaum {\it et al.} 
(NLO and NNLO, respectively).\protect\cite{EGM98}
The notation for the empirical points is the same as in
Fig.~2.}
\end{figure}

\begin{figure}[t]
\vspace*{-2.0cm}
\hspace*{-1.7cm}
\psfig{figure=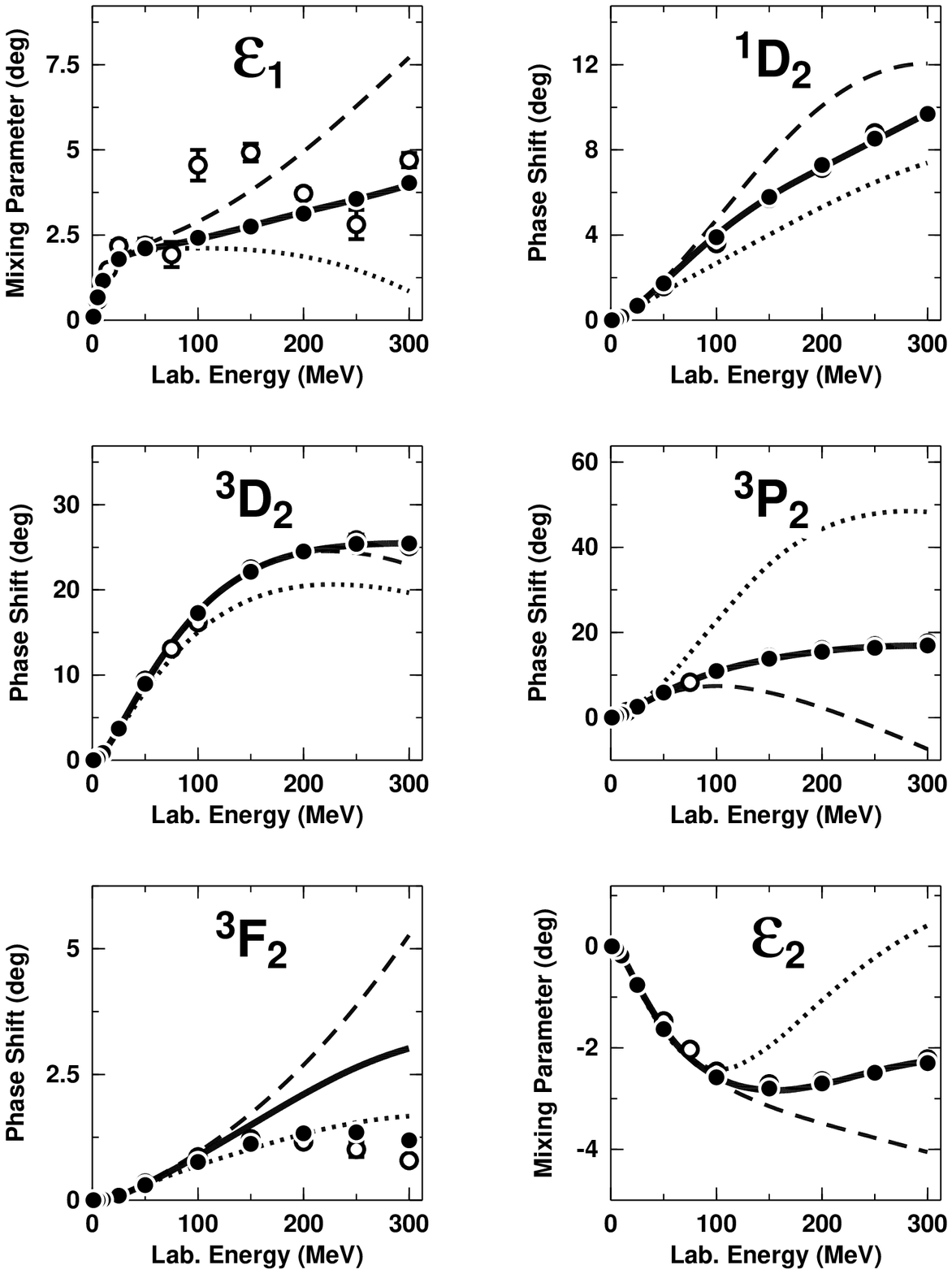,height=18.0cm}
\vspace*{-3.3cm}
\caption{Mixing angle $\epsilon_1$ and $J=2$ phase parameters.
Notation as in Fig.~4.}
\end{figure}

In Figs.~4 and 5, we show the phase shifts of neutron-proton ($np$) scattering
for lab.\ energies below 300 MeV and partial waves with $J\leq 2$.
The solid line represents the result from the chiral NN potential developed in
the present work. {\it The reproduction of the empirical phase shifts by our model
is excellent.} For comparison, we also show the phase shift predictions
by two chiral models recently developed by 
Epelbaum {\it et al.}\cite{EGM98} (dotted and dashed curves in Figs.~4 and 5).
In the upper part of Table~1, we give our results for the effective range parameters
of the $S$ waves which agree accurately with the empirical values.
We note that our present chiral potential is charge-independent and adjusted
to the $np$ data.

\subsection{The Deuteron}

\begin{table}
\caption{Two- and three-nucleon low-energy data.}
\footnotesize
\begin{tabular}{llllll}
\hline
\hline
 & Idaho-A$^a$ & Idaho-B$^a$ & CD-Bonn\cite{Mac01} & AV18\cite{WSS95} & Empirical$^b$ \\
\hline
\hline
\multicolumn{2}{l}{\bf Low-energy $np$ scattering} \\
$^1S_0$ scattering length (fm) &-23.75&-23.75&-23.74&-23.73&-23.74(2)\\
$^1S_0$ effective range  (fm) &2.70&2.70&2.67&2.70&2.77(5)\\
$^3S_1$ scattering length (fm) &5.417&5.417&5.420&5.419&5.419(7)\\
$^3S_1$ effective range (fm) &1.750&1.750&1.751&1.753&1.753(8)\\
\hline
{\bf Deuteron properties}\\
Binding energy (MeV) &
2.224575& 2.224575&
 2.224575 & 2.224575 & 2.224575(9) \\
Asympt.\ $S$ state (fm$^{-1/2}$) &
0.8846 &0.8846 
& 0.8846& 0.8850 & 0.8846(9) \\
Asympt. $D/S$ state         & 
0.0256 & 0.0255 &
0.0256& 0.0250&0.0256(4)\\
Deuteron radius (fm)   &
1.9756$^c$ &
1.9758$^c$ &
 1.970$^c$ &
 1.971$^c$ &
 1.9754(9)$^d$ \\
Quadrupole moment (fm$^2$) &
0.281$^e$ &
0.284$^e$ &
 0.280$^e$ & 
0.280$^e$ &
 0.2859(3) \\
$D$-state probability (\%)    & 
4.17 &
4.94 &
4.85 & 5.76  \\
\hline
{\bf Triton binding} (MeV) &8.14&8.02&8.00&7.62&8.48\\
\hline
\hline
\end{tabular}
\indent
$^a$Chiral NN potential of the present work.\\
$^b$For references concerning the empirical data, 
see Tables XIV and XVIII of Ref.\cite{Mac01}.\\
$^c$With meson-exchange current (MEC) and relativistic corrections.\cite{FMS97}\\
$^d$Reference\cite{Hub98}.\\
$^e$Including MEC and relativistic corrections in the amount of 0.010 fm$^2$.\cite{Hen95}
\end{table}

The reproduction of the deuteron parameters is shown in the middle part of Table~1.
We present results for two versions of our chiral NN potential, dubbed 
`Idaho-A' and `Idaho-B'.\footnote{We note that the phase shifts 
represented by the solid line in Figs~4 and 5 are for Idaho-B; 
however, the ones for Idaho-A are so close to Idaho-B
that they could not be distinguished on the scale of the figure.}
The main difference between the two models is in the $D$-state probability
of the deuteron, $P_D$. Even though $P_D$ is not an observable, it is
of theoretical interest since the binding energies of few- and many-nucleon
systems depend on it (cf.\ triton results at the bottom of Table~1).
As mentioned before, the predictive power of the chiral model is very
limited and it is possible to construct chiral potentials that fit
the $^3S_1$, $^3D_1$, and $\epsilon_1$ phase parameters 
up to 300 MeV and the empirical deuteron properties accurately, but 
have $D$-state probabilities that range from 3 to 6\%.
Such a large variation of $P_D$ is not possible within the meson model
of nuclear forces.

Remarkable are the results produced by our chiral potentials
for the deuteron radius which agree accurately with the latest
empirical value obtained by using the isotope-shift method.\cite{Hub98} 
All NN potentials of the past 
(Table 1 includes two representative examples,
namely, CD-Bonn\cite{Mac01} and AV18\cite{WSS95})
fail to reproduce this very precise new value for the deuteron radius.\cite{FMS97}
Our chiral NN potentials are the first to predict this value right.

In Fig.~6, we display the deuteron wave functions derived from
our chiral potential (Idaho-B) by the solid line and compare them
with wave functions based upon conventional NN potentials from the
recent past. Characteristic differences are noticeable; in particular,
the chiral wave functions are shifted towards larger $r$ which explains the
larger deuteron radius.

Concerning 
the triton binding energy predictions given at the bottom of Table~1,
we like to comment that the results 
for Idaho-A and B are obtained in a
34-channel Faddeev calculation with no charge-dependence (i.~e., using the $np$
potential throughout), while the corresponding calculations with CD-Bonn
and AV18 take charge-denpendence into account.

\begin{figure}[t]
\vspace*{-2.0cm}
\hspace*{0.8cm}
\psfig{figure=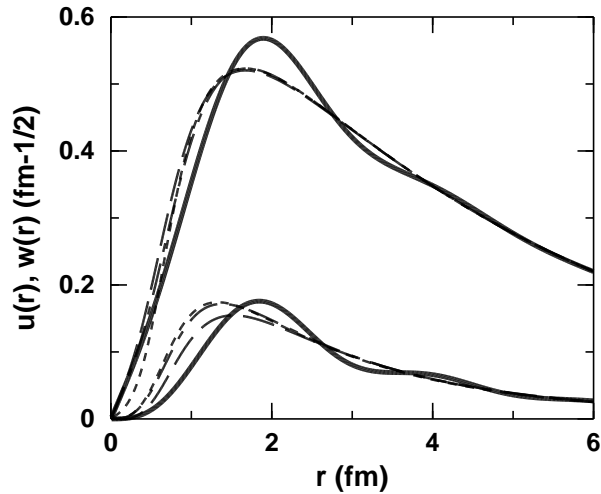,height=12.0cm}
\vspace*{-3.0cm}
\caption{Deuteron wave functions: the larger curves are $S$-waves, 
the smaller ones $D$-waves. The solid line represents the wave functions derived
from our chiral NN potential (Idaho-B). The dashed, dash-dotted, and dotted lines
refer to the wave functions of the CD-Bonn\protect\cite{Mac01}, 
Nijm-I\protect\cite{Sto94}, and AV18\protect\cite{WSS95} potentials, respectively.}
\end{figure}

\section{Summary and Conclusions}

We have constructed an {\it accurate} chiral NN potential.
The model includes one- and two-pion exchange contributions
up to chiral order three and contact terms (which represent
the short range force) up to order four.
Within this framework, the NN phase shifts below 300 MeV lab.\
energy and the properties of the deuteron
are reproduced with high-precision.

Due to the very quantitative nature of this new chiral NN potential,
it represents a reliable and promising starting point
for exact few-body calculations and
microscopic nuclear many-body theory.

\section*{Acknowledgments}
We gratefully acknowledge useful discussions with B. van Kolck,
E. Epelbaum, W. Gl\"ockle, N. Kaiser, U. Mei\ss ner, and M. Robilotta. 
This work was supported in part by the U.S. National Science
Foundation under Grant No.~PHY-0099444 and by the Ram\~on Areces
Foundation (Spain).

\end{document}